\documentclass[conference]{IEEEtran}

\IEEEoverridecommandlockouts
\usepackage[utf8]{inputenc}

\usepackage{geometry}
\def\skeleton{0}
\def\review{0} 
\def\arxiv{1}
\def\withAck{1}
\usepackage[cmex10]{amsmath}
\usepackage{amssymb}
\usepackage{mathtools}
\usepackage{bm}
\usepackage{algorithmic}
\usepackage{graphicx}
\usepackage[dvipsnames]{xcolor}
\usepackage[acronym,toc]{glossaries}
\usepackage{cleveref}
\usepackage{siunitx}
\usepackage{tabularx}
\usepackage{booktabs}
\usepackage{pgfplots, tikz}
\usepackage{pgfplotstable}
\pgfplotsset{compat=1.18}
\usepgfplotslibrary{groupplots}
\usepackage{ifthen}
\usetikzlibrary{arrows.meta, positioning, external}
\usepackage{layouts}
\usepackage{environ}
\usepackage{placeins}
\usepackage{subcaption}
\usepackage[caption=false, font=footnotesize]{subfig}
\usepackage{stfloats}
\usepackage{textcase}
\usepackage[tablename=TABLE,font=small]{caption}
\usepackage{xcolor}
\usepackage{upgreek}
\usepackage{balance}
\usepackage{titlesec}
\usepackage{float}
\usepackage{times}
\usepackage[cmintegrals]{newtxmath}
\usepackage[inline, shortlabels]{enumitem}

\if\review1
\usepackage{todonotes}
\else
\usepackage[disable]{todonotes}
\fi

\usepackage[activate={true,nocompatibility},
    final,
    tracking=true,
    kerning=true,
    spacing=true,
    factor=1100,
    stretch=25,
    shrink=25,
    letterspace=-30
]{microtype}

\makeatletter
\newsavebox{\measure@tikzpicture}
\NewEnviron{scaletikzpicturetowidth}[1]{%
    \def\tikz@width{#1}%
    \begin{lrbox}{\measure@tikzpicture}%
        \BODY
    \end{lrbox}%
    \pgfmathparse{#1/\wd\measure@tikzpicture}%
    \BODY
}
\makeatother

\newacronym{3gpp}{3GPP}{3rd Generation Partnership Project}
\newacronym{5g}{5G}{fifth generation}
\newacronym{6g}{6G}{sixth generation}

\newacronym{admm}{ADMM}{Alternating Directions of Multipliers Method}
\newacronym{anm}{ANM}{Atomic Norm Minimization}
\newacronym{adc}{ADC}{Analog-to-Digital Converter}
\newacronym{awgn}{AWGN}{Additive White Gaussian Noise}
\newacronym{asic}{ASIC}{Application Specific Integrated Circuit}
\newacronym{arpack}{ARPACK}{ARnoldi PACKage}
\newacronym{api}{API}{Application Programmable Interface}
\newacronym{aut}{AUT}{Antenna Under Test}
\newacronym[plural=AOI, firstplural=Areas of Interest (AOI)]{aoi}{AOI}{Area of Interest}
\newacronym{ai}{AI}{Artifical Intelligence}
\newacronym{aic}{AIC}{Akaike Information Criterion}

\newacronym{bp}{BP}{Basis Pursuit}
\newacronym{bpdn}{BPDN}{Basis Pursuit Denoising}
\newacronym{blue}{BLUE}{Best Linear Unbiased Estimator}

\newacronym{cnn}{CNN}{Convolutional Neural Network}
\newacronym{crb}{CRB}{Cramér-Rao Bound}
\newacronym{cs}{CS}{Compressed Sensing}
\newacronym{csi}{CSI}{channel state information}
\newacronym{cf}{CF}{Crest factor}
\newacronym{cr}{CR}{Compression Ratio}
\newacronym{cmos}{CMOS}{Complementary Metal Oxide Semiconductor}
\newacronym{cots}{COTS}{Commercial-Off-The-Shelf}
\newacronym{cp}{CP}{cyclic prefix}
\newacronym{cpu}{CPU}{Central Processing Unit}
\newacronym{cfar}{CFAR}{Constant False Alarm Rate}
\newacronym{ca-cfar}{CA-CFAR}{Cell-Averaging Constant False Alarm Rate}
\newacronym{osca-cfar}{OS-CA-CFAR}{Ordered-Statistic \glslink{ca-cfar}{Cell-Averaging CFAR}}
\newacronym{csca-cfar}{CS-CA-CFAR}{Censored Statistics \glslink{ca-cfar}{Cell-Averaging CFAR}}
\newacronym{so-cfar}{SO-CA-CFAR}{Smallest-Of \glslink{ca-cfar}{Cell-Averaging CFAR}}
\newacronym{cut}{CUT}{cell under test}
\newacronym{cvx}{CVX}{Convex Optimization toolboX}

\newacronym{dac}{DAC}{Digital-to-Analog Converter}
\newacronym{doa}{DoA}{Direction of Arrival}
\newacronym{dod}{DoD}{Direction of Departure}
\newacronym[plural=DMC,firstplural=Dense Multipath Components (DMC)]{dmc}{DMC}{Dense Multipath Components}
\newacronym{das}{DAS}{Delay-and-Sum}
\newacronym{dft}{DFT}{Discrete Fourier Transform}
\newacronym{idft}{IDFT}{Inverse Discrete Fourier Transform}
\newacronym{dct}{DCT}{Discrete Cosine Transform}
\newacronym{dlearn}{DL}{Deep Learning}
\newacronym{dl}{DL}{downlink}
\newacronym{dsp}{DSP}{Digital Signal Processor}
\newacronym{dnn}{DNN}{Deep Neural Network}
\newacronym{dpss}{DPSS}{Discrete Prolate Spheroidal Sequences}

\newacronym{eadf}{EADF}{Effective Aperture Distribution Function}
\newacronym{etadf}{ETADF}{Effective Time-Aperture Distribution Function}
\newacronym{esprit}{ESPRIT}{Estimation of Signal Parameters via Rotational Invariance Techniques}
\newacronym{ett}{ETT}{Eigenvalue Threshold Test}
\newacronym{edc}{EDC}{Efficient Detection Criterion}
\newacronym{eft}{EFT}{Exponential Fitting Test}

\newacronym{fa}{FA}{false alarm}
\newacronym{fc}{FC}{fully-connected}
\newacronym{fht}{FHT}{Fast Hadamard Transform}
\newacronym[longplural={Fast Fourier Transforms}]{fft}{FFT}{Fast Fourier Transform}
\newacronym{fmcw}{FMCW}{Frequency-Modulated Continuous-Wave}
\newacronym{fpga}{FPGA}{Field Programmable Gate Array}
\newacronym{fri}{FRI}{Finite Rate of Innovation}
\newacronym{fim}{FIM}{Fisher Information Matrix}
\newacronym{fmc}{FMC}{Full Matrix Capture}
\newacronym{fn}{FN}{false negative}
\newacronym{fista}{FISTA}{Fast Iterative Shrinkage-Thresholding Algorithm}
\newacronym{fp}{FP}{false positive}
\newacronym{fr2}{FR2}{Frequency Range 2}
\newacronym{frvm}{FRVM}{Fast Relevance Vector Machine}

\newacronym{gan}{GAN}{Generative Adversarial Network}
\newacronym{gfcs}{grid-free CS}{grid-free compressive sensing}
\newacronym{gnb}{gNB}{gNodeB}
\newacronym{gm}{GM}{Gaussian mixture}
\newacronym{gpu}{GPU}{Graphical Processing Unit}
\newacronym{gtd}{GTD}{Geometrical Theory of Diffraction}

\newacronym{hrpe}{HRPE}{High Resolution Parameter Estimator}
\newacronym{hfss}{Ansys HFSS}{Ansys High Frequency Electromagnetic Simulation Software}

\newacronym[longplural={Inverse Fast Fourier Transforms}]{ifft}{IFFT}{Inverse Fast Fourier Transform}
\newacronym{ir}{IR}{Impulse Response}
\newacronym{iid}{iid}{independent and identically distributed}
\newacronym{iir}{IIR}{Infinite Impulse Response}
\newacronym{irf}{IRF}{Impulse Response Function}
\newacronym{isac}{ISAC}{Integrated Sensing and Communications}
\newacronym{ista}{ISTA}{Iterative Shrinkage-Thresholding Algorithm}

\newacronym{kf}{KF}{Kalman Filter}
\newacronym{kld}{KLD}{Kullback-Leibler Divergence}

\newacronym{ls}{LS}{Least Squares}
\newacronym{lasso}{LASSO}{Least Absolute Shrinkage and Selection Operator}
\newacronym{lse}{LSE}{Line Spectral Estimation}
\newacronym{lfsr}{LFSR}{Linear Feedback Shift Register}
\newacronym{lo}{LO}{Local Oscillator}
\newacronym{los}{LOS}{line-of-sight}
\newacronym{lti}{LTI}{Linear Time-Invariant}
\newacronym{lam}{LAM}{Large Area Monitoring}

\newacronym{mimo}{MIMO}{multiple input multiple output}
\newacronym{mmv}{MMV}{multiple measurement vectors}
\newacronym{ml}{ML}{maximum likelihood}
\newacronym{mmse}{MMSE}{misspecified mean squared error}
\newacronym{mmwave}{mmWave}{millimeter wave}
\newacronym{mse}{MSE}{mean squared error}
\newacronym{bce}{BCE}{Binary Crossentropy}
\newacronym{mlbs}{MLBS}{Maximum Length Binary Sequence}
\newacronym{mo}{MO}{Modulation Order}
\newacronym{mpc}{MPC}{Multipath Component}
\newacronym{msm}{MSM}{M-Sequence Method}
\newacronym{mwc}{MWC}{Modulated Wideband Converter}
\newacronym{mpm}{MPM}{Matrix Pencil Method}
\newacronym{mpu}{MPU}{Microprocessor Unit}
\newacronym{mri}{MRI}{Magnetic resonance imaging}
\newacronym{music}{MUSIC}{Multiple Signal Classification}
\newacronym{mkl}{MKL}{Math Kernel Library}
\newacronym{mcrb}{MCRB}{Misspecified Cramér-Rao Bound}
\newacronym{mmle}{MMLE}{Misspecified Maximum-Likelihood Estimator}

\newacronym{ndt}{NDT}{Nondestructive Testing}
\newacronym{nde}{NDE}{Nondestructive Evaluation}
\newacronym{nlos}{NLOS}{non-line-of-sight}
\newacronym{nneigh}{NN}{Nearest Neighbor}
\newacronym{nist}{NIST}{National Institute of Standards and Technology}

\newacronym{ofdm}{OFDM}{orthogonal frequency-division multiplexing}
\newacronym{omp}{OMP}{Orthogonal Matching Pursuit}
\newacronym{oop}{OOP}{Object Oriented Programming}

\newacronym{pa}{PA}{power amplifier}
\newacronym{pdf}{pdf}{probability density function}
\newacronym{pdp}{PDP}{Power Delay Profile}
\newacronym{phd}{PHD}{probability hypothesis density}
\newacronym{pn}{PN}{Pseudo-Noise}
\newacronym{poc}{PoC}{Proof of Concept}
\newacronym{pwc}{PWC}{Plane Wave Compounding}
\newacronym{pwi}{PWI}{Plane Wave Imaging}
\newacronym{pura}{PURA}{Patch Uniform Rectangular Array}
\newacronym{pymax}{PyMAX}{Python Maximization Approach}
\newacronym{psf}{PSF}{Point Spread Function}
\newacronym{pts}{PTS}{Pseudo-True Solution}

\newacronym{qpsk}{QPSK}{Quadrature phase-shift keying}

\newacronym{ram}{RAM}{Random Access Memory}
\newacronym{rc}{RC}{Raised Cosine}
\newacronym{rcs}{RCS}{radar cross section}
\newacronym{rd}{RD}{Random Demodulator}
\newacronym{relu}{ReLU}{Rectified Linear Unit}
\newacronym{rem}{REM}{Reconstruction Error Metric}
\newacronym{resnet}{ResNet}{Residual Neural Network}
\newacronym{ric}{RIC}{Restricted Isometry Constant}
\newacronym{rimax}{RIMAX}{Richter Maximization Approach}
\newacronym{rf}{RF}{radio frequency}
\newacronym{rip}{RIP}{Restricted Isometry Property}
\newacronym{rms}{RMS}{root mean squared}
\newacronym{rmse}{RMSE}{Root Mean Squared Error}
\newacronym{roc}{ROC}{Receiver Operating Characteristic}
\newacronym{roi}{ROI}{Region of Interest}
\newacronym{rvm}{RVM}{Relevance Vector Machine}
\newacronym{ru}{RU}{Radio Unit}
\newacronym{rx}{RX}{receiver}

\newacronym{scf}{SCF}{spatial correlation function}
\newacronym{samurai}{SAMURAI}{Synthetic Aperture Measurements of Uncertainty in Angle of Incidence}
\newacronym[plural=SC,firstplural=Specular Components (SC)]{sc}{SC}{Specular Components}
\newacronym{sdp}{SDP}{semi-definite program}
\newacronym{sdr}{SDR}{Signal to Diffuse Ratio}
\newacronym{simd}{SIMD}{Single Instruction Multiple Data}
\newacronym{svd}{SVD}{singular value decomposition}
\newacronym{svm}{SVM}{Support Vector Machine}
\newacronym{soe}{SOE}{Sparsity Order Estimation}
\newacronym{sgd}{SGD}{Stochastic Gradient Descent}
\newacronym{stuca}{StUCA}{Stacked Uniform Circular Array}
\newacronym{spucpa}{SPUCPA}{Stacked Polarimetric Uniform Circular Patch Array}
\newacronym{suca}{SUCA}{Stacked Uniform Circular Array}
\newacronym{saft}{SAFT}{Synthetic Aperture Focusing Technique}
\newacronym{sota}{SOTA}{State of the Art}
\newacronym{ssd}{SSD}{Solid State Device}
\newacronym{ssr}{SSR}{Sparse Signal Recovery}
\newacronym{sa}{SA}{Synthetic Aperture}
\newacronym{spw}{SPW}{Single Plane Wave}
\newacronym{shm}{SHM}{Structural Health Monitoring}
\newacronym{snr}{SNR}{Signal-to-Noise Ratio}
\newacronym{srrc}{SRRC}{Square-root Raised Cosine}
\newacronym{stela}{STELA}{Soft-Thresholding with Exact Line Search Algorithm}
\newacronym{siso}{SISO}{Single Input Single Output}
\newacronym{simo}{SIMO}{Single Input Multiple Output}
\newacronym{swe}{SWE}{Spherical Wave Expansion}
\newacronym{sme}{SME}{Spherical Mode Expansion}
\newacronym{sage}{SAGE}{Space-Alternating Generalized Expectation-Maximization}

\newacronym{tdd}{TDD}{time division duplex}
\newacronym{th}{T\&H}{Track and Hold}
\newacronym{tf}{TF}{Transfer Function}
\newacronym{tof}{ToF}{Time-of-Flight}
\newacronym{toc}{TOC}{Tracker Operating Characteristic}
\newacronym{tn}{TN}{true negative}
\newacronym{tp}{TP}{true positive}
\newacronym{twista}{TWISTA}{Two-step Iterative Shrinkage-Thresholding Algorithm}
\newacronym{tx}{TX}{transmitter}

\newacronym{uca}{UCA}{uniform circular array}
\newacronym{ue}{UE}{User Equipment}
\newacronym{ura}{URA}{Uniform Rectangular Array}
\newacronym{ul}{UL}{uplink}
\newacronym{ula}{ULA}{Uniform Linear Array}
\newacronym{uwb}{UWB}{Ultra-Wideband}
\newacronym{usndt}{US-NDT}{Ultrasonic Non-destructive Testing}

\newacronym{vna}{VNA}{Vector Network Analyser}
\newacronym{vsh}{VSH}{Vector Spherical Harmonics}
\def\BibTeX{{\rm B\kern-.05em{\sc i\kern-.025em b}\kern-.08em
T\kern-.1667em\lower.7ex\hbox{E}\kern-.125emX}}

\newcommand{\eg}{e.g.,\ }

\newcommand{\iec}{i.\,e., }

\newcommand\blfootnote[1]{%
  \begingroup
  \renewcommand\thefootnote{}\footnote{#1}%
  \addtocounter{footnote}{-1}%
  \endgroup
}

\definecolor{QC1}{HTML}{0077BB}
\definecolor{QC2}{HTML}{33BBEE}
\definecolor{QC3}{HTML}{009988}
\definecolor{QC4}{HTML}{EE7733}
\definecolor{QC5}{HTML}{CC3311}
\definecolor{QC6}{HTML}{EE3377}
\definecolor{QC7}{HTML}{BBBBBB}
\definecolor{QCW}{HTML}{000000}

\ifx\qty\undefined
    \let\oldnum\num
    \renewcommand{\num}[2][ignored]{\oldnum{#2}}

    \ifx\qty\undefined
        \newcommand{\qty}[3][ignored]{\SI{#2}{#3}}
    \fi

    \ifx\qtyrange\undefined
        \newcommand{\qtyrange}[4][ignored]{\SIrange{#2}{#3}{#4}}
    \fi
\fi

\usepackage{}

\begin{document}

\title{Reliable Non-Line-of-Sight Intrusion Detection with Integrated Sensing and Communications Hardware
}

\author{\IEEEauthorblockN{
        Paolo Tosi\IEEEauthorrefmark{1}\IEEEauthorrefmark{2},
        Maximilian Bauhofer\IEEEauthorrefmark{3},
        Marcus Henninger\IEEEauthorrefmark{1},
        Laurent Schmalen\IEEEauthorrefmark{2},
        and Silvio Mandelli\IEEEauthorrefmark{1}}
\IEEEauthorblockA{
    \IEEEauthorrefmark{1}Nokia Bell Labs Stuttgart \;\;
    \IEEEauthorrefmark{2}Karlsruhe Institute of Technology (KIT) \;\;
    \IEEEauthorrefmark{3}University of Stuttgart, Germany}\; 
    E-mail: paolo.tosi@nokia.com
    }
\maketitle

\begin{abstract}
\Gls{nlos} sensing has the potential to enable use cases like intrusion detection in occluded areas, increasing the value provided by \gls{isac} in future 6G cellular networks.
In this paper, we present a reliable \gls{nlos} intrusion detection system based on a millimeter-wave \gls{isac} proof-of-concept. 
By leveraging reflections off a large surface, the proposed system addresses the challenge of detecting moving targets in cluttered indoor industrial scenarios where the direct line-of-sight is obstructed. 
A signal processing pipeline including a tracking stage, comparing a \gls{kf}- and a \gls{phd} filter-based approach, is applied to detect targets and track movements in \gls{nlos}. Experimental validation conducted in the ARENA2036 industrial research campus demonstrates that our system can reliably detect target presence in \gls{nlos} while avoiding false alarms. Tests with synthetically generated false peaks further demonstrate the robustness of our system to false alarms.
Overall, the results underline the potential of \gls{nlos} \gls{isac} as a promising technology for enabling intrusion detection and monitoring use cases.

\glsresetall
\end{abstract}

\if\arxiv1
\blfootnote{This work has been submitted to the IEEE for possible publication. Copyright may be transferred without notice, after which this version may no longer be accessible.}
\else
\vspace{0.2cm}
\fi

\section{Introduction}
\if\skeleton1
\color{blue}
\begin{itemize}
    \item ISAC brings radar functionalities to 6G networks
    \item For certain use cases, e.g. indoor or urban scenarios the signal may reach the target only via multipath propagation (NLOS)
    \item Opportunities of NLOS sensing compared to cameras, LiDAR 
    \item Literature on NLOS radar feasibility
    \item We define the problem as a track initiation problem in low-SNR and cluttered scenarios
    \item Literature on NLOS track initiation
    \item we compare our Kalman filter- based approaches
\end{itemize}
\color{black}
\fi

The upcoming \gls{6g} of cellular networks promises to introduce the capability to extract information about the environment, essentially operating the network as a radar.
This joint operation of radar and communication services is commonly referred to as \gls{isac} \cite{Mandelli_Henn_Bau_Wild_2023}.
Recent efforts by the \gls{3gpp} have focused on determining the feasibility of \gls{isac}, as well as to define use cases for realistic deployments~\cite{3gpp_22837}.
One of the most interesting applications of \gls{isac} is intrusion detection, for example for highway and railway safety, intersection monitoring and smart home monitoring~\cite{9829746}.
Furthermore, smart factory floors were identified as another significant area of interest, \eg for detection and tracking of automated vehicles and human personnel in production plants or warehouses \cite{Ghosh_2025}.
Such environments are often inherently rich in multipath components and obstacles creating difficult propagation conditions.
Targets may not be in \gls{los} of the system, causing the radio signal to reach those objects only via multiple reflections \eg off walls or large reflectors. To still offer sensing capabilities in such scenarios, \gls{rf}-based \gls{nlos} sensing can be performed, whereas other technologies (e.g., cameras or LiDARs) have no or very limited \gls{nlos} capabilities. 
Exploiting the multipath components of radio signals can improve sensing coverage in dense areas, reducing the need to densify the network.

\Gls{nlos} sensing experiments with dedicated radar equipment have already been reported in the literature. 
For instance, behind-the-corner vehicular detection for safe intersection monitoring exploiting a passive reflector has been presented in \cite{Solomitckii_NLOS_w_reflector1, Solomitckii_NLOS_w_reflector2}. 
There, a large planar surface, \eg a wall from surrounding buildings or a large metallic surface, was used as a reflector to relay the radar signal around the obstacles, improving the effective visibility of the system.
The authors of \cite{Gustafsson_urban_micro_doppler} used an experimental coherent high-resolution X-band radar to track and extract micro-Doppler features of walking people in \gls{nlos} conditions by exploiting buildings as reflecting surfaces in urban intersection scenarios.
In \cite{Li_Nlos_breathing}, a frequency-modulated continuous-wave millimeter wave signal was used to extract detailed features about human targets in \gls{nlos}, such as their breathing rate.
The aforementioned studies showed that \gls{nlos} detection is a promising approach for improving the effective coverage of radar systems in cases of limited visibility and highly cluttered scenarios.
However, most of them were performed using dedicated radar equipment that does not exhibit the limitations often encountered with \gls{isac}, which is typically based on systems primarily designed for communications purposes.
For instance, waveform as well as frame structure and numerology are bound to the communication standards. 
One example is the \gls{tdd} transmission pattern, which results in undesirable impulsive sidelobes in the ambiguity function of targets in radar images.
Our previous work~\cite{Feasibility_Nlos} explored processing schemes for detecting \gls{nlos} targets from multipath propagation in an indoor industrial scenario with \gls{isac} hardware.
This work highlighted the general feasibility of \gls{nlos} sensing using commercially available communication hardware.
The observed target, however, was not completely in \gls{nlos} conditions, as a direct \gls{los} path was still present due to the transmitted beam's sidelobe.
Moreover, a robust detection and tracking scheme enabling reliable intrusion detection has not been implemented.

\begin{figure*}[!t]
    \centering
    \vspace{0.12cm}
    \resizebox{0.8\textwidth}{!}{
\def\nodetextsize{\scriptsize}
\begin{tikzpicture}[
    block/.style={
        draw,
        rectangle,
        minimum height=0.7cm,
        minimum width=0.5cm,
        align=center,
        font=\scriptsize
    },
    rounded corners=3pt,
    tx/.style={block, fill=green!20},
    channel/.style={block, fill=green!20},
    rx/.style={block, fill=green!20},
    c_est/.style={block, fill=orange!20},
    det/.style={block, fill=orange!20},
    trak/.style={block, fill=cyan!20},
    arrow/.style={-latex, line width=0.5pt}
]


\node[tx] (mod) {OFDM\\Transmitter};
\node[channel, right=1.8cm of mod] (channel) {Wireless\\Channel};
\node[rx,  right=1.2cm of channel] (rec) {OFDM\\Receiver};
\node[c_est, right=1.4cm of rec] (c_est) {CSI Matrix\\Computation};

\node[det, below=0.3cm of c_est] (IDFT) {IDFT $N\!\uparrow$ \\ DFT $M\!\!\rightarrow$};
\node[det, below=0.3cm of rec] (cfar) {CA-CFAR\\Detector};
\node[det, below=0.3cm of channel] (peak_conf) {TDD Peak\\Detection};
\node[trak, below=0.3cm of mod] (trak) {Tracking\\Filter};

\draw[arrow] (mod) -- node[above, font=\nodetextsize, xshift=-8pt]{$\mathbf{X}[n,m]$} (channel);
\draw[arrow] (channel) -- (rec);
\draw[arrow] (rec) -- node[above, font=\nodetextsize]{$\mathbf{Y}[n,m]$}(c_est);

\draw [arrow] ($(mod.east)+(1.3,0)$) node[circle,fill=black,inner sep=1.0pt]{} 
            -- ++ (0.0,0.6) -| node [pos=0.99] {} (c_est);
\draw[arrow]
    (c_est.east) -- node[above, font=\nodetextsize, rotate=0, xshift=8pt, yshift=-0pt]{$\mathbf{H}[n,m]$}($(c_est.east)+(0.5,0.0)$)  -- ($(IDFT.east)+(0.6, -0.0)$) -- (IDFT.east);

\draw[arrow] (IDFT.west) -- node[above, font=\nodetextsize]{$\mathbf{S}[k,l]$}(cfar.east);
\draw[arrow] (cfar.west) -- (peak_conf.east);
\draw[arrow] (peak_conf.west) --  node[above, font=\nodetextsize]{Peaks $Z$}(trak.east);

\end{tikzpicture}
    }
    \caption{Proposed signal processing pipeline. Operations performed by the \gls{5g}-compliant communication hardware are highlighted in green, while orange boxes represent the sensing processing steps and the blue box the target tracking process.
    }
    \label{fig:signal_pipeline}
\end{figure*}
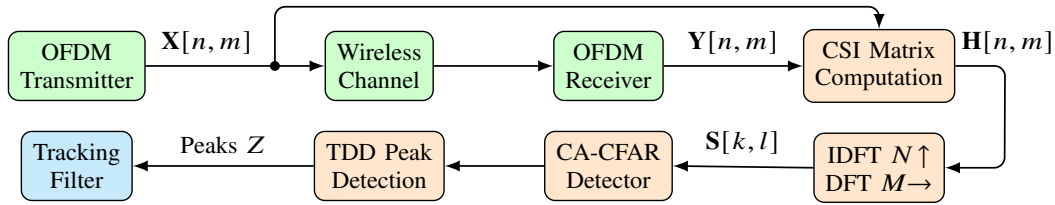 

In this work, we expand the intrusion detection use case applied to indoor industrial floors by recording a fully \gls{nlos} indoor scenario.
We provide a demonstration of of the first \gls{isac} \gls{nlos} intrusion detection system with tracking using the  \gls{poc} described in~\cite{wild2023integrated}.
Our objective is to detect the presence of a moving target in an obscured area leveraging the multipath reflections off a nearby wall.
To that end, we collect measurements with a single human target in \gls{nlos} moving at different speeds in the ARENA~2036 industrial research campus. Moreover, we implement a signal processing pipeline comparing a \gls{kf}-based track confirmation system and a state-of-the-art \gls{phd} filter to showcase reliable \gls{nlos} intrusion detection capability.

\section{ISAC system setup}
\if\skeleton1
\color{Green}
\begin{itemize}
    \item OFDM radar system
    \item CSI estimated via element-wise division of reflected and transmitted radio frames
    \item radar image obtained by computing the 2D periodogram via DFT
    \item target plots are obtained using the TDD peak detection
\end{itemize}
\color{black}
\fi

\subsection{System description}
\label{sec:sysModel_description}

In this work, we used measurements from our \gls{isac} \gls{poc}, which uses commercially available 5G communication hardware and operates in \gls{fr2} at central frequency $\qty{27.4}{\giga\hertz}$ \cite{wild2023integrated}.
The system comprises a standard \gls{gnb} \gls{ru}, operating as \gls{tx}, and a Sniffer \gls{ru} operating in \gls{ul} mode as \gls{rx}.
The \glspl{ru} are quasi co-located and synchronized, allowing to treat the system as a monostatic sensing setup.
The \glspl{ru} consists of 12 antenna elements per row and 8 antenna elements per column and can transmit with 2 polarizations.
The system is equipped with analog beamforming, which allows the selection of a fixed beam from a predefined beam codebook.
The \gls{gnb} transmits \gls{5g}-compliant \gls{ofdm} radio frames with $\qty{10}{\ms}$ duration in \gls{tdd} mode using numerology $\mu = 3$~\cite{3gpp_38211}. 
Each \gls{tdd} pattern extends over $T_{\text{TDD}} = 1.25$~ms and comprises $M_{\text{DL}} = 104$ \gls{dl} and $M_{\text{UL}} = 36$ \gls{ul} symbols, \iec a \gls{dl}/\gls{ul} ratio of approx. 3:1.
Each radio frame~$\mathbf{X} \in \mathbb{C}^{N\times M}$ consists of~$M$ \gls{ofdm} symbols with $N$ subcarriers, spaced by $\Delta f$, carrying complex-modulated symbols. 
The signal interacts with the environment by illuminating objects and reflecting off them.

\subsection{Sensing processing}
\label{sec:sysModel_targetEst}

A server receives the IQ samples from \gls{gnb} and Sniffer on a per-frame basis, and computes the \gls{csi} matrix $\mathbf{H} \in \mathbb{C}^{N\times M}$ via element-wise division of the  received frame $\mathbf{Y} \in \mathbb{C}^{N\times M}$ by the (known) transmitted one
\begin{align}
    \mathbf{H}[n,m] = \frac{\mathbf{Y}[n,m]}{\mathbf{X}[n,m]}.
\end{align}
The \gls{csi} matrix contains the contributions due to reflections from $P$ objects in the environment, each of which is positioned~$r_p$ meters from the system and moving with relative radial velocity~$v_p$ to the system.
After obtaining $\mathbf{H}$, the range-Doppler periodogram (radar image) $\mathbf{S}$ is computed by performing a \gls{dft} over the \gls{ofdm} symbols and an \gls{idft} over the subcarriers
\vspace{-0.2cm}
\begin{equation}
\begin{split}
\mathbf{S}[k,l] &= \frac{1}{N'M'}\left|\sum_{k=0}^{N'} \Biggl(\sum_{l=0}^{M'} \mathbf{H}[k, l]e^{-j2\pi\frac{lm}{M'}}\Biggr)e^{j2\pi\frac{kn}{N'}} \right|^2, 
\end{split}
\label{eq:periodogram}
\end{equation}
where $N' = 2^{\lceil{\log_2{N}}\rceil}$ and $M' = 2^{\lceil{\log_2{M}}\rceil}$ are the number of rows and columns of $\mathbf{H}$ after zero padding.

Reflections due to the $P$ objects lead to peaks in the radar image.
To determine whether a bin of the radar image is relevant, \iec corresponds to a peak, we use a \gls{ca-cfar} detector \cite{Richards_Scheer_Holm_2010}.
From the coordinates of the peaks in the radar image, one can determine the parameters associated with them (e.g., distance, relative speed, radar cross section). 
The \gls{tdd} transmission pattern implies that \gls{ul} symbols cannot be used for sensing \cite{3gpp_38211}, resulting in spectral holes in the time domain.
These holes create spectral replicas of peaks, spaced by the speed resolution multiplied by the number of \gls{tdd} patterns within the observation aperture as discussed in~\cite{Henninger_TDD}.
To avoid false alarms due to this effect, we employ a \gls{tdd} peak detection routine exploiting knowledge of the time domain windowing to extract a set of peaks $Z$ from each image, while rejecting the spectral replicas. 
For more details regarding this procedure, please refer to \cite{Henninger_TDD}.
Each peak consists of a range and relative speed estimate $z_p = [\hat{r}_p, \hat{v}_p]$ possibly corresponding to targets.

Fig. \ref{fig:signal_pipeline} presents a full scheme of the communication and radar signal processing pipeline.

\subsection{Scenario Description}
\label{subsec:scenario}

\begin{figure*}[!t]
    \centering
    \vspace{0.2cm}
    \begin{subfigure}[t]{0.23\textwidth}
      \centering
      \input{images/poc/poc_photo_highlighted.tikz}
      \caption{\gls{gnb} (top) and Sniffer (bottom).}
    \end{subfigure}
    \hfill
    \begin{subfigure}[t]{0.32\textwidth}
      \centering
      \includegraphics[height=4.2cm, keepaspectratio]{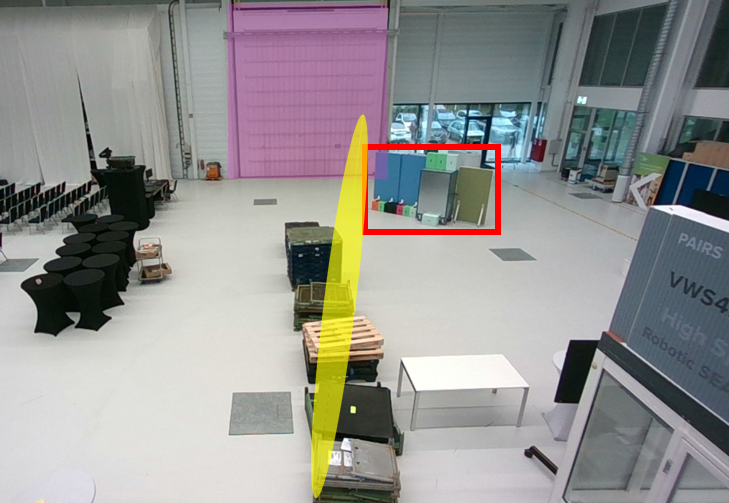}
      \caption{Camera view of the test scene.}
      \label{fig:arena_camera}
    \end{subfigure}
    \hfill
    \begin{subfigure}[t]{0.32\textwidth}
      \centering
      \includegraphics[height=4.2cm, keepaspectratio]{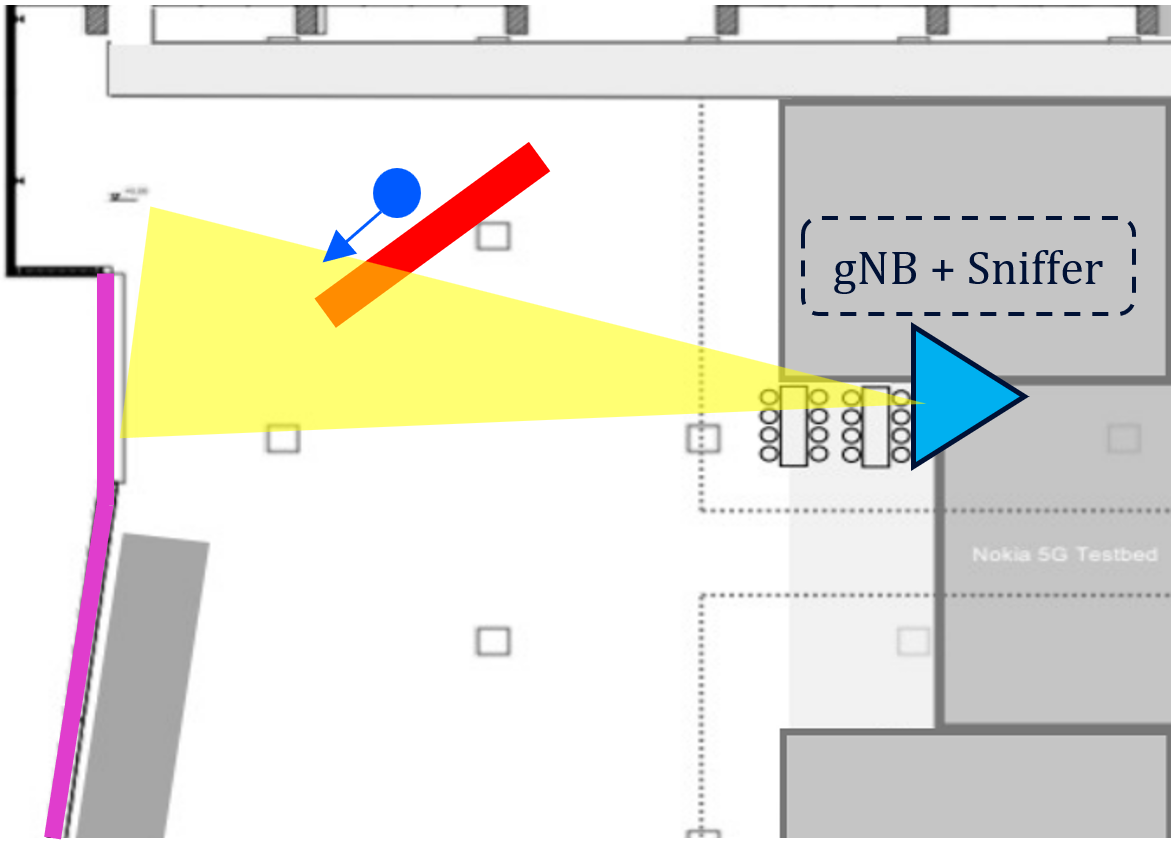}
      \caption{Map of the test area.}
    \end{subfigure}
    \caption{Measurement scenario in the ARENA2036. The distance between the monostatic sensing setup and the cargo gate (highlighted in pink) is $\approx\qty{23}{\m}$. The target moves behind the blocker wall (red box), while the transmitted beam reaches the surveilled area via single-bounce reflection off the cargo gate.
    }
    \label{fig:arena_setup_full}
\end{figure*}

\begin{figure}[t]
    \vspace{-0.2cm}
    \def\targetRectColor{Magenta}

\begin{tikzpicture}
 
    \begin{axis}[
        axis on top,
        height=7.5cm,
        width=8.5cm,
        enlargelimits=false,
        xmin=-5, xmax=5,
        xlabel near ticks,
        ymin=0, ymax=45,
        ytick={0,20,...,60},
        ylabel near ticks,
        xlabel={Rel. Velocity [m/s]},
        ylabel={Range [m]},
        label style={font=\small},
        tick label style={font=\small},
        legend style={font=\footnotesize},
        legend cell align={left},
        legend entries={Peak, Static reflector},
        legend style=
        	{fill=white, 
        	fill opacity=0.4, 
        	draw opacity=1, 
        	text opacity=1, 
        	nodes={scale=1, transform shape}, 
        	at={(0.97,0.05)}, 
        	anchor=south east,
            }
        ]

         \addplot[forget plot] graphics[xmin=-5.0, xmax=5.0, ymin=0, ymax=45, ] {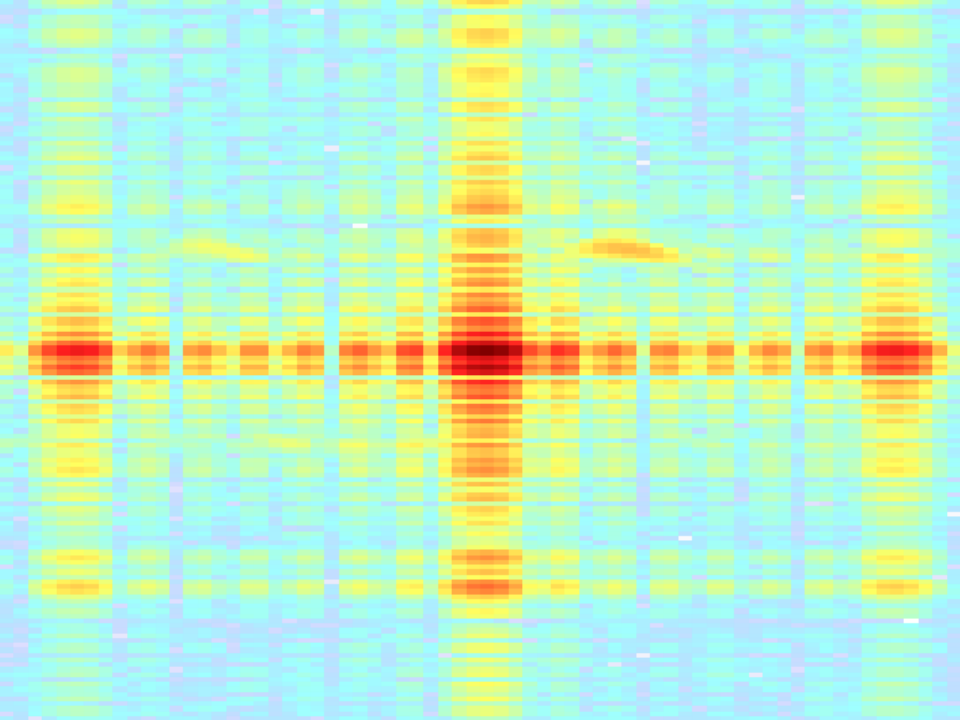};
         
         \draw[red, dashed,  line width=1pt] (axis cs:-4.45,24) rectangle (axis cs:4.45,44);
         \draw[MidnightBlue, dashdotted,  line width=1pt] (axis cs:-4.85,0.5) rectangle (axis cs:-3.65,44.5);
         \draw[MidnightBlue, dashdotted,  line width=1pt] (axis cs:4.85,0.5) rectangle (axis cs:3.65,44.5);
         \node[font=\footnotesize, text=MidnightBlue] at (axis cs:-4.25,1.6) {TDD};
         \draw[\targetRectColor, solid,  line width=1pt] (axis cs:-0.7,21.5) rectangle (axis cs:0.7,24.5);
         \node[font=\small, text=red] at (axis cs:2.8,42.5) {NLOS};
         
         \addplot[
            only marks,
            mark=x,
            mark size=3pt,
            color=ProcessBlue,
            line width=1pt,
            forget plot,
        ] table[
            x index=1,  
            y index=0,  
            col sep=comma,
        ] {Data/periodogram/target_plots.txt};

        \addlegendimage{only marks, mark=x, color=ProcessBlue, 
          mark size=3pt,                 
          line width=1pt}

        \addlegendimage{only marks, mark=square, color=\targetRectColor, 
          mark size=3pt,                 
          line width=1pt}
      
    \end{axis}
 
\end{tikzpicture}
 
    \caption{Example output of the TDD peak detection step, before discarding static peaks. The blue rectangles highlight the \gls{tdd} replicas. The strongest static target (pink box) is the main reflector. The wall blocking the \gls{los} path is detected at $[\qty{15.7}{\m}, \qty{0.01}{\m/\s}]$. The moving target in \gls{nlos} is detected at $[\qty{29.3}{\m}, \qty{1.67}{\m/\s}]$.}
    \vspace{-0.1cm}
    \label{fig:periodogram}
\end{figure}
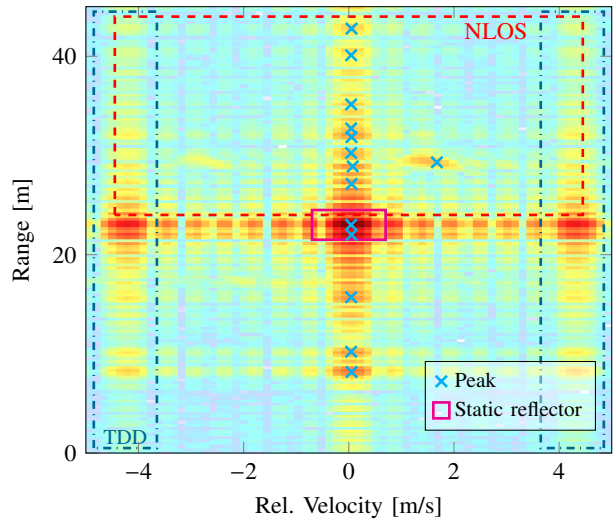

We emulated a \gls{nlos} scenario in the ARENA2036 industrial research campus in Stuttgart, Germany. 
An illustration of the setup is shown in Fig. \ref{fig:arena_setup_full}. 
Furthermore, Fig.~\ref{fig:periodogram} presents a periodogram from this environment showing detected peaks as output of the sensing processing described previously.

To ensure \gls{nlos} coverage, the signal must be directed towards a large obstacle with a sufficient \gls{rcs}, allowing it to reflect towards the target in the \gls{nlos} area. 
In the absence of precise knowledge of the environment in which the system is deployed, the \gls{nlos} region can be defined by selecting a strong static reflector beyond which no \gls{los} component can occur. 
Consequently, the \gls{nlos} area encompasses all ranges beyond the reflector's range (red box in Fig.~\ref{fig:periodogram}).

A calibration measurement can be conducted in the absence of moving targets within the test scenario, thereby extracting information regarding the static targets in the environment, including the main reflector defining the \gls{nlos} area.
In our case, the objects in the red rectangle in Fig.~\ref{fig:arena_camera} are used to block the direct \gls{los} component of the signal transmitted by the \gls{isac} equipment. 
The cargo gate at the back (highlighted in pink) serves as the main reflector causing the strongest target return in the radar image (pink box in Fig. \ref{fig:periodogram}).
The camera view (Fig. \ref{fig:arena_camera}), which is co-located with the \gls{gnb}, shows that the target is not directly visible, \iec no \gls{los} component exists. 
The surveilled \gls{nlos} area is illuminated by the single-bounce propagation path with the beam orientation kept fixed.
The transmitted beam has a main lobe width of $\ang{14}$, which is enough to illuminate a significant portion of the cargo gate and create a reflection onto the surveilled area.

For the intrusion detection use case, it is sufficient to determine the presence of an intruder by detecting movement in the surveilled area.
For this reason, we discard peaks with zero Doppler for \gls{nlos} processing retaining only the subset $Z_{\text{mov}}$.
In the example periodogram of Fig. \ref{fig:periodogram}, only the target peak $\mathbf{z}_p = [\qty{29.31}{\m}, \qty{1.67}{\m/\s}]$ is retained.
The list of non-static peaks is passed to the tracking stage described in the next Section.

\section{NLOS Target Tracking}
To distinguish false detections in our noisy environment from the presence of actual targets, we incorporate a tracking step for robust target detection.
The tracker leverages the previous measurements/states to obtain an estimate of the actual, current state of the system.
We compare two approaches:
\begin{enumerate*}[(A)]
    \item a \gls{kf}-based approach with with nearest-neighbor data association, gating, and a covariance-based track confirmation test; and
    \item a \acrfull{phd} filter from the class of random finite set trackers \cite{vo_gmphd}. 
\end{enumerate*}
The KF approach is conceptually simpler but requires auxiliary logic to handle track initiation and termination, whereas the PHD filter provides these capabilities natively at the cost of not maintaining labelled tracks over time.
\subsection{Kalman Filter}
The confirmation system maintains a list of tentative tracks~$\mathcal{T}$.
Each track $\tau \in \mathcal{T}$ is represented by a state estimate \mbox{$\mathbf{x}_k^\tau = [r_k^\tau, v_k^\tau]$}, a state covariance matrix~$\mathbf{P}_k^\tau$ and a track timer~$t$.
When the track list is empty, tracks are initialized upon receiving valid peaks from the detection step.

At each time step, tentative tracks evolve their state $\mathbf{x}_{k|k-1}^\tau$ following the \gls{kf} prediction equation. 
We assume that the signal reflected by the target will not exhibit large accelerations, especially in the evaluated indoor scenario.
Therefore, we consider a constant-velocity state transition matrix
\begin{equation}
    \boldsymbol{F}_{k-1} = 
    \begin{bmatrix}
            1 & -\Delta t \\
            0 & 1
\end{bmatrix},
\end{equation}
where $\Delta t$ is the time update interval.
The negative sign accounts for the range-rate direction reversal introduced by the single-bounce reflection of the signal off the wall, whereas a positive sign is used for \gls{los} targets. 

Upon receiving new peaks $Z_{\text{mov},k}$, only the closest one $\mathbf{z}_k^\tau$ is used for the update of each tentative track.
The list of peaks associated with active tracks is $\hat{Z}_k$.
This results, in practice, in a nearest-neighbor assignment approach.
This approach suffices for our single-target scenario where few peaks arrive per update, and more complex approaches (e.g., Global nearest-neighbor) would achieve a similar result.
Furthermore, peaks associated with each track are gated before the update step, i.e., are discarded if their normalized distance from the track estimate in the state space exceeds a the gating threshold $\tilde{d}$.
If no peak is associated to a track, the track is advanced through the state prediction only.
Each track is updated following the \gls{kf} update equation and by increasing the track timer.
Remaining detected peaks, that were not associated to any of the tentative tracks, are used to initialize new tracks.

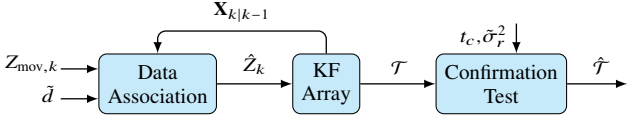
\begin{figure}[!t]
    \centering
    \resizebox{0.99\columnwidth}{!}{
\def\nodetextsize{\scriptsize}
\begin{tikzpicture}[
    block/.style={
        draw,
        rectangle,
        minimum height=0.8cm,
        minimum width=0.5cm,
        align=center,
        font=\footnotesize
    },
    rounded corners=3pt,
    tx/.style={block, fill=green!20},
    channel/.style={block, fill=green!20},
    rx/.style={block, fill=green!20},
    c_est/.style={block,},
    det/.style={block, fill=orange!20},
    trak/.style={block, fill=cyan!20},
    arrow/.style={-latex, line width=0.5pt},
]

\node[trak, ] (data_a) {Data\\Association};
\node[trak, right=1.0cm of data_a] (KF) {KF\\Array};
\node[trak, right=1.0cm of KF] (trak) {Confirmation\\Test};

\draw[arrow] ($(data_a.west)+(-0.5,0.2)$) --  node[above, font=\nodetextsize, xshift=-18pt, yshift=-5pt, align=center]{$Z_{\text{mov},k}$}($(data_a.west)+(-0.0,0.2)$);
\draw[arrow] ($(data_a.west)+(-0.5,-0.2)$) --  node[above, font=\nodetextsize, xshift=-12pt, yshift=-5pt, align=center]{$\tilde{d}$}($(data_a.west)+(-0.0,-0.2)$);
\draw[arrow] (data_a.east) --  node[above, font=\nodetextsize, xshift=-0pt, align=center]{$\hat{Z}_k$}(KF.west);
\draw[arrow] (KF.east) --  node[above, font=\nodetextsize, align=center]{$\mathcal{T}$}(trak.west);
\draw[arrow]
    (KF.north) -- ($(KF.north)+(0.0,0.3)$) -- node[above, font=\nodetextsize, align=center]{$\mathbf{X}_{k|k-1}$}($(data_a.north)+(0.0,0.3)$) -- (data_a.north);
\draw[arrow]
    ($(trak.north)+(0.2,0.4)$) -- node[left, font=\nodetextsize, yshift=1pt]{$t_c$,$\tilde{\sigma}_r^2$}($(trak.north)+(0.2,0.0)$);
\draw[arrow]
    ($(trak.east)+(0.0,0.0)$) -- node[above, font=\nodetextsize, align=center, xshift=1.0pt]{$\mathcal{\hat{T}}$}($(trak.east)+(0.8,0.0)$);

\end{tikzpicture}
    }
    \caption{\gls{kf}-based track confirmation system. Incoming peaks ${Z}_{\text{mov},k}$ are associated to tentative tracks $\mathcal{T}$ via nearest-neighbor assignment with gating threshold $\tilde{d}$.
    Each track is updated through the \gls{kf} and checked against the confirmation criteria: minimum duration $t_c$ and maximum range covariance $\tilde{\sigma}^2_r$. 
    Confirmed tracks $\mathcal{\hat{T}}$ trigger an intrusion alarm.}
    \label{fig:confirmation_diagram}
\end{figure}

After all tentative tracks have been advanced, the system performs a confirmation check to determine if any of the tentative tracks should be added to the set of confirmed tracks~$\mathcal{\hat{T}}$ or deleted.
Two thresholds $\tilde{\sigma}_r^2$ and~$t_c$ are defined, which represent the maximum range covariance threshold and the minimum tracking time, respectively.
A track is confirmed if its \gls{kf} remained active at least for a time $t_c$ and its range covariance never exceeded $\tilde{\sigma}_r^2$ during this period.
The track is discarded if $\tilde{\sigma}_r^2$ is exceeded.
Fig. \ref{fig:confirmation_diagram} shows a diagram summarizing the steps of the track confirmation process.


\subsection{PHD Filter}
The \gls{phd} filter avoids the direct association by a weighted combination of every observation $Z_{\text{mov},k}$ with all states, and handles target birth and death directly -- in contrast to the previous \gls{kf} proposal.
This filter is unlabelled, i.e., it does not provide a continuous track in time, but it estimates the set of target states at each time step.
Specifically, we use the \gls{gm} implementation.
The filter propagates a multi-target state intensity function~$\nu$, represented as a \gls{gm}~\cite{vo_gmphd}.
The intensity~$\nu$ approximates the expected number of targets at any candidate state $\boldsymbol{x}$ in the scenario.
This allows for a closed-form solution under the prerequisite of Gaussian distributed noise and a linear prediction and observation model, which is suitable for our setup, just like for the \gls{kf}.
The \gls{gm} for the posterior intensity $\nu_{k-1}(\boldsymbol{x})$ at time $k-1$ is given by
\begin{equation}
    \nu_{k-1}(\boldsymbol{x}) = \sum_{i=1}^{J_{k-1}}w_{k-1}^{(i)}\mathcal{N}\left( \boldsymbol{x}; \boldsymbol{m}_{k-1}^{(i)},\boldsymbol{P}_{k-1}^{(i)} \right)\,,
\end{equation}
with $J$ \gls{gm} components with weights $w^{(i)}$, means $\boldsymbol{m}^{(i)}$, and covariances $\boldsymbol{P}^{(i)}$.

In the prediction step, the posterior weights are decreased by the probability of survival $p_{S,k}$ -- enabling track death -- and then predicted to the current time using the state transition matrix $\boldsymbol{F}_{k-1}$ and process noise covariance $\boldsymbol{Q}_{k-1}$ with
{\setlength{\jot}{-2pt}
\begin{multline}
    \nu_{S,k|k-1}(\boldsymbol{x}) = p_{S,k}\sum_{j=1}^{J_{k-1}}w_{k-1}^{(j)}\mathcal{N}\left( \boldsymbol{x};\boldsymbol{F}_{k-1}\boldsymbol{m}_{k-1}^{(j)},\right.\\
    \left.\boldsymbol{Q}_{k-1}+\boldsymbol{F}_{k-1}\boldsymbol{P}_{k-1}^{(j)}\boldsymbol{F}_{k-1}^\intercal \right) \,.
\end{multline}
These survival components $\nu_{S,k|k-1}(\boldsymbol{x})$ are paired with newly born components, themselves a \gls{gm} and enabling track birth, forming the predicted intensity
\begin{equation}
    \nu_{k|k-1}(\boldsymbol{x}) = \nu_{S,k|k-1}(\boldsymbol{x}) + \gamma_k(\boldsymbol{x}) \,.
\end{equation}
Note that spawning components $\nu_{\beta,k|k-1}(\boldsymbol{x})$ could also be added.

In the update step, the measurements $\boldsymbol{z} \in Z_{\text{mov},k}$ are used in the components $\nu_{D,k} (\boldsymbol{x};\boldsymbol{z})$, updating $\nu_{k|k-1}(\boldsymbol{x})$, for the updated intensity
\begin{equation}
    \nu_{k}(\boldsymbol{x}) = (1-p_{D,k})\nu_{k|k-1}(\boldsymbol{x}) + \sum_{\boldsymbol{z} \in Z_{\text{mov},k}}\nu_{D,k}(\boldsymbol{x};\boldsymbol{z})\,,
\end{equation}
with probability of detection $p_{D,k}$.
After the update, the growing number of \gls{gm} components need pruning and merging.
The states correspond to the highest-weighted components of the \gls{gm}, where the number of selected components $\hat{P}$ is determined by the total sum of weights.

\section{Experimental validation}
\subsection{Data collection}
\label{subsec:data_collection}

Based on the setup presented in Section \ref{subsec:scenario}, we conducted several experiments involving a single human target.
The parameters of the \gls{isac} system and the tracking filters are summarized in Table~\ref{tab:meas_params}. 
Moreover, to further test the \gls{fa} rate of our system in less ideal conditions, we generated synthetic false detections to feed into the tracking step.

\begin{table}[!b]
    \centering
    \caption{Measurement parameters}
    \label{tab:meas_params}
    \begin{tabularx}{0.95\linewidth}{Xp{2.5cm}}
        \toprule
        Parameter                            &   Value      \\ \midrule
        \textbf{ISAC System} \\
        Carrier frequency $f_c$              &   27.4 GHz                         \\
        Number of subcarriers $N$            &   1584                             \\
        Subcarrier spacing $\Delta f$        &   120 kHz                          \\
        Total bandwidth                      &   $N\Delta f = 190$ MHz            \\
        Num. of OFDM symbols $M$             &   1120                             \\ 
        OFDM symbol time $T$ (incl. CP)        & $8.92 \;\upmu$s                    \\ 
        \midrule
        \textbf{Kalman Filter} \\
        Gating norm. distance $\tilde{d}$                 & 5.0                    \\
        Min. tracking time $t_c$            & $\qty{60}{\milli\s}$               \\
        Range process noise standard deviation & $\qty{3.5}{\m}$       \\
        Speed process noise standard deviation & $\qty{1.8}{\meter/\second}$       \\
        Range measurement noise standard deviation & $\qty{1.8}{\m}$   \\
        Speed measurement noise standard deviation & $\qty{1.8}{\meter/\second}$        \\
        \midrule
        \textbf{PHD Filter} \\
        Survival probability                 & $p_{S} = 0.98$                    \\
        Detection probability                & $p_D = 0.9$                       \\
        State transition matrix              & $\boldsymbol{F}_{k-1} = \bigl( \begin{smallmatrix}1 & -\Delta t\\ 0 & 1\end{smallmatrix}\bigr)$ \\
        Birth weight                         & $w_B = 1 \times 10^{-5}$ \\
        \gls{gm} merging radius              & $0.7$ \\
        \gls{gm} pruning max number of components              & $30$ \\
        \bottomrule
    \end{tabularx}
\end{table}

\paragraph{True target scenario}
The measurement was recorded while the target paced back and forth behind the blocker and we ensured that it was not visible from the camera stream or from any LOS component.
The experiment was replicated by having the subject walk and run in \gls{nlos}, with a total of two trials.
The different pacing speeds between experiments are used to test detection conditions, exploiting the Doppler information to separate moving targets from static clutter.
These experiments constitute a \gls{tp} set, in which the target is almost always in motion, except for brief changes of direction.
To convey a better understanding of the underlying scenario, Fig.~\ref{fig:kf_full_meas} shows the range estimates from a single \gls{kf} and the \gls{phd} filter applied to the recorded dataset of a person running in \gls{nlos}.
In this example, a target trajectory from the \gls{nlos} signal can be established while the person moves with sufficient speed.
The back and forth motion of the walking person can be clearly observed from both the detections and the range estimate.
A low detection rate can be observed when the target changes direction, as it is static for a short time.
In the interval $\qty{2.5}{\s} - \qty{3.5}{\s}$, where the tracker does not receive target information, the cardinality $\hat{P}$ of the \gls{phd} filter is zero and no state estimate is stored, while the \gls{kf} is able to retain the track, but its estimation covariance grows.

\paragraph{Empty scenario}
\label{par:empty_scenario}
By recording the scenario without moving targets, we collected a \gls{tn} measurement set, where we expect the system to not detect the presence of any target.

\paragraph{Synthetic false detections set}
To test the system performance in even more challenging conditions, we expand the measurement set by generating synthetic false peaks as additional outputs of the target estimation step.
For our scenario, we modeled the number of false peaks per frame as a Poisson-distributed random variable with expected rate $\lambda$.
Each false detection consists of range and speed values, sampled from a uniform distribution defined over the same  \gls{nlos} measurement space of the real radar images.

\subsection{Validation Methodology}
Each dataset consists of a list of \gls{csi} matrices, recorded at $T_f = \qty{10}{\milli\s}$ intervals over a total observation time $T_m \approx \qty{10}{\s}$.

In practical scenarios, however, the observation window is likely shorter and alarms should ideally be raised in under a second.
\begin{figure}[!t]
    \vspace{-0.1cm}
    \def\datafile{Data/PHD/range_est.txt}
\def\detectionsfile{Data/kf_data/detections.txt}
\def\datafileKF{Data/kf_data/range_est.txt}
\def\imagewidth{0.97}

\def\colorOne{BlueGreen}
\def\colorTwo{MidnightBlue}
\def\colorThree{Bittersweet}

\def\figHeight{5.5cm}
\def\figWidth{8.7cm}

\begin{tikzpicture}

        \begin{axis}[
            unbounded coords=jump,
			height = \figHeight,
    		width=\figWidth,
            line width = 0.5pt,
    		xlabel={Time (s)},
    		ylabel={Range est. (m)},
    		xmax = 8.1,
            xmin =  0.0,
    		ymin=28.5,
		    ymax = 34.5,
            ylabel style={black, font=\footnotesize,at={(axis description cs:-0.05,.5)},rotate=0,anchor=south},
		    enlargelimits = false,
    		xmajorgrids=true,
            ymajorgrids=true,
            grid style={solid, opacity=0.6},
            label style={font=\small},
            ylabel style={\colorTwo, font=\footnotesize},
            tick label style={font=\footnotesize},
            yticklabel style={\colorTwo, font=\footnotesize},
		    legend style={font=\footnotesize,
                fill=white, 
            	fill opacity=0.4, 
            	draw opacity=1, 
            	text opacity=1, 
                legend columns = 3,
            	nodes={scale=1, transform shape}, 
            	at={(0.5,1.16)},   
            	anchor=north,
            },
            clip mode=individual, 
		]

        \addplot[
            forget plot,
            \colorOne,
            only marks,
            mark=x,
            mark size=0.7pt,
        ] table[
            x index=0,
            y expr=\thisrowno{1} - 3,
            col sep=comma,
        ] {\detectionsfile};

         \addplot[
   	    color=\colorTwo,
        mark options={solid},
        forget plot,
        mark=none,
        line width=1.3pt,
        on layer=main, 
        ]
    	plot table[x expr=\thisrowno{0}, y expr=\thisrowno{1} - 3] {\datafile};

        \addplot[
        dashed,
   	    color=\colorTwo,
        mark options={solid},
        forget plot,
        mark=none,
        line width=1.3pt,
        on layer=main, 
        ]
    	plot table[x expr=\thisrowno{0}, y expr=\thisrowno{1} - 3] {\datafileKF};

        \addlegendimage{gray, solid, line width=1.2pt}
        \addlegendimage{gray, dashed, line width=1.2pt}
        \addlegendimage{\colorOne, only marks, mark=x, mark size=3pt, line width=1.2pt}

        \legend{
            PHD filter,
            KF,
            Peaks,
        }
        
        \end{axis}

        \begin{axis}[
            axis y line*=right,
            unbounded coords=discard,
			height = \figHeight,
    		width=\figWidth,
            line width = 0.5pt,
    		ylabel={Range cov. $\sigma_r^2$ ($\si{\m}^2$)},
    		xmax = 8.1,
            xmin =  0.0,
    		ymin=-0.05,
		    ymax = +3.0,
            ylabel style={Bittersweet, font=\footnotesize},
            ylabel shift = -4 pt,
		    enlargelimits = false,
    		xmajorgrids=true,
            ymajorgrids=true,
            grid style={dashdotted, opacity=0.6},
            label style={font=\small},
            tick label style={font=\footnotesize},
            yticklabel style={Bittersweet, font=\footnotesize},
		]

        \addplot[
        densely dashed,
   	    color=Bittersweet,
        mark options={solid},
        forget plot,
        smooth,
        tension=0.7,
        mark=none,
        line width=1pt]
    	plot table[x expr=\thisrowno{0}, y expr=\thisrowno{1}] {Data/kf_data/range_cov.txt};
        
        \end{axis}

    \end{tikzpicture}
    \caption{Top: PHD and KF range estimates (blue lines) for the full dataset of a running person in \gls{nlos}. When the target stops and changes direction, the system does not generate peaks (light blue marker). In that case, the estimated number of targets is zero and the PHD estimate is interrupted. The KF retains the track but its covariance (bottom orange line) grows during measurement gaps.
    }
    \label{fig:kf_full_meas}
\end{figure}
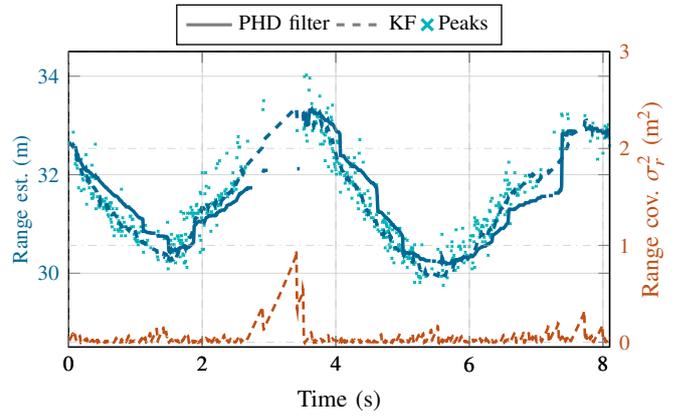
To increase the number of data points, we therefore split the recorded data into sub-measurements of length $\tilde{T}_m$, each overlapping the previous one by $\tilde{T}_m - T_f$, \iec shifting by one frame. For each sub-measurement, the presence of an intruder is checked.
Radar images are generated from the \gls{csi} matrices on a per-radio frame basis and a list of peaks is produced as the output of the detector step.
In intrusion detection scenarios, we assume that a target must move to enter the surveilled area. Hence, static detections are discarded, as mentioned in Sec.~\ref{subsec:scenario}.
The tracking process applied to each sub-measurement acts as a binary classifier deciding whether a target is present or not during this time.
To obtain the \mbox{\gls{roc}} curve of our system, we define the \gls{tp} rate as the ratio between the number of sub-measurements from the \gls{tp} set resulting in \textit{intruder detected} and the total number of \gls{tp} sub-measurements.
Here, we refer to sub-measurements in which any of the \gls{kf} tracks is confirmed or the \gls{phd} filter estimates a number of moving targets $\hat{P}_{\text{mov}}\geq1$ as \textit{intruder detected}.
Similarly, the \gls{fp} rate is obtained using the sub-measurements from the  \gls{fp} set.

\subsection{Results}

By discarding peaks corresponding to static targets, no target detection was generated from the empty \gls{tn} scenario described in Sec. \ref{subsec:scenario}.
Consequently, our system was able to avoid raising any false alarm throughout the whole dataset, while reliably detectiong the moving intruder.

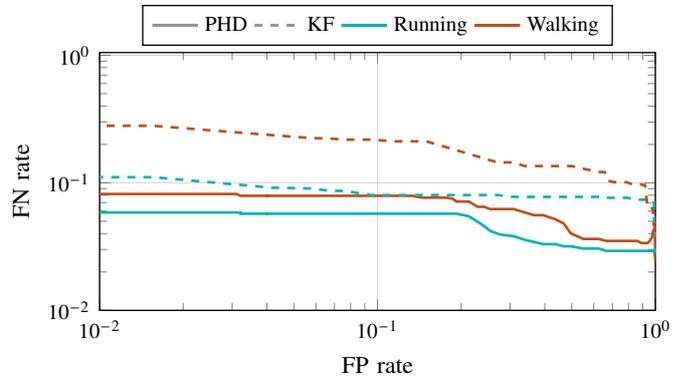
\begin{figure}[t!]
    \vspace{0.0cm}
    \centering
    \def\filenamePhd{phd_roc_300ms.txt}
\def\filenamePhdOne{phd_roc_100ms.txt}
\def\filenamePhdPointEight{phd_roc_80ms.txt}
\def\scale{1}
\def\plotPhd{1}

\def\colorOne{BlueGreen}
\def\colorTwo{Bittersweet}
\def\colorThree{MidnightBlue}

\begin{tikzpicture}
        \begin{loglogaxis}[
			height = 5cm,
    		width=1\columnwidth,
            line width = 0.5pt,
    		xlabel={FP rate},
    		ylabel={FN rate},
    		xmax = 1.0,
            xmin =  0.01,
    		ymin=0.01,
		    ymax = 1.05,
            ylabel style={font=\footnotesize,at={(axis description cs:.-0.15,.5)},rotate=0,anchor=south},
            xlabel near ticks, 
            ylabel near ticks,
		    enlargelimits = false,
    		xmajorgrids=true,
            ymajorgrids=true,
            grid style={solid, opacity=0.6},
            label style={font=\small},
            tick label style={font=\footnotesize},
    		legend pos = south west,
		    legend style={font=\footnotesize,
                fill=none, 
            	fill opacity=0.2, 
            	draw opacity=1, 
            	text opacity=1, 
                legend columns = 4,
                legend cell align={center},
            	scale = 1,
                at={(0.5,1.185)},   
            	anchor=north,
            },
		]

        \addplot[
   	    color=\colorOne,
        solid,
        mark options={solid},
        line width=1pt,
        forget plot]
    	plot table[x expr=\thisrowno{0}, y expr=1- \thisrowno{1}] {Data/PHD/\filenamePhd};

        \addplot[
   	    color=\colorTwo,
        solid,
        mark options={solid},
        line width=1pt,
        forget plot]
    	plot table[x expr=\thisrowno{0}, y expr=1 - \thisrowno{2}] {Data/PHD/\filenamePhd};

        \addplot[
   	    color=\colorOne,
        dashed,
        mark options={solid},
        line width=1pt,
        forget plot]
    	plot table[x expr=\thisrowno{0}, y expr=1- \thisrowno{1}] {Data/kf_roc/kf_roc_300ms.txt};

        \addplot[
   	    color=\colorTwo,
        dashed,
        mark options={solid},
        line width=1pt,
        forget plot]
    	plot table[x expr=\thisrowno{0}, y expr=1 - \thisrowno{2}] {Data/kf_roc/kf_roc_300ms.txt};
        
        \addlegendimage{Gray, solid, line width=1.2pt}
        \addlegendentry{PHD}

        \addlegendimage{Gray, dashed, line width=1.2pt}
        \addlegendentry{KF}
        
        \addlegendimage{\colorOne, solid, line width=1.2pt}
        \addlegendentry{Running}

        \addlegendimage{\colorTwo, solid, line width=1.2pt}
        \addlegendentry{Walking}


        \end{loglogaxis}
\end{tikzpicture}
    \vspace{-0.5cm}
    \caption{ Detection performance (FN rate vs. FP rate) for \mbox{$\tilde{T}_m = 300$~ms}. Running target in light blue; walking target in orange. Solid lines: PHD filter
 (swept over birth weight $w_B$); dashed lines: KF (swept over covariance threshold $\tilde{\sigma}^2_r$). Lower-left is better. The running target achieves superior separation from clutter due to its higher Doppler.
    }
    \label{fig:res_roc_walking_running}
\end{figure}

To further stress the system in presence of noise or other distortions, we test the detection on the synthetic false detection dataset, which models false detections as presented in Sec.~\ref{subsec:data_collection}.
The average number of false detections per image is Poisson-distributed with expected rate $\lambda=0.8$.
Fig. \ref{fig:res_roc_walking_running} shows the \gls{roc} curve for \gls{phd}- and \gls{kf}-based target tracking solutions, 
for walking and running scenarios with sub-measurement duration $\tilde{T}_m = \qty{300}{\milli\s}$ 
for an increasing value of the weight $w_B$ of the birth density~$\gamma_k$ of the \glspl{gm} of the \gls{phd} filter and of the range covariance threshold $\tilde{\sigma}_r^2$ for the \gls{kf}.
The \gls{fn} (missed detection) rate is defined as $1 - \text{\gls{tp}}$.
One can observe that the higher speed of the running target results in better separation from static clutter, and, consequently, leads to a better detection performance compared to the walking measurement.

As illustrated in Fig.~\ref{fig:res_roc_obs_time}, a comparison is made of the detection performance of the two approaches for sub-measurement durations of $\qty{80}{\milli\s}$, $\qty{100}{\milli\s}$ and $\qty{300}{\milli\s}$. This is achieved by combining the walking and running datasets into a single dataset.
The target presence is detected for all measurements, except those in which the target stops to change direction. This is due to the fact that detections are filtered by ignoring zero-Doppler detections.
The \gls{phd}-based approach demonstrates superior performance compared to the \gls{kf}-based one.
Given the measurement time of $\tilde{T}_m = \qty{300}{\milli\s}$, the \gls{phd} filter is capable of detecting the presence of a target in $\qty{96.75}{\percent}$ of sub-measurements. In comparison, the performance of the \gls{kf} saturates at $\qty{95.5}{\percent}$.
As expected, the missed detection (\gls{fn}) rate increases for a smaller sub-measurement length.
This result indicates that, in similar scenarios, an intruder can be identified within a similar time frame using few updates of the filter. 
Note that only a single detection is sufficient to raise an alarm, which almost guarantees the detection of an intruder during the overall observation time $T_m$.

\begin{figure}[t]
    \centering    \centerline{\hspace{1.4cm}\pgfplotslegendfromname{legend:roc_phd_kf}\hfill}
    \vspace{-0.10cm}
    \def\filenamePhd{phd_roc_300ms.txt}
\def\filenamePhdOne{phd_roc_100ms.txt}
\def\filenamePhdPointEight{phd_roc_80ms.txt}
\def\scale{1}
\def\plotPhd{1}

\def\colorOne{BlueGreen}
\def\colorTwo{Bittersweet}
\def\colorThree{MidnightBlue}

\begin{tikzpicture}
        \begin{loglogaxis}[
			height = 4.85cm,
    		width=1\columnwidth,
            line width = 0.5pt,
    		xlabel={FP rate},
    		ylabel={FN rate},
    		xmax = 1.0,
            xmin =  0.01,
    		ymin=0.01,
		    ymax = 1.05,
            ylabel style={font=\footnotesize,at={(axis description cs:.-0.15,.5)},rotate=0,anchor=south},
            xlabel near ticks, 
            ylabel near ticks,
		    enlargelimits = false,
    		xmajorgrids=true,
            ymajorgrids=true,
            grid style={solid, opacity=0.6},
            label style={font=\small},
            tick label style={font=\footnotesize},
    		legend pos = south west,
		    legend style={font=\footnotesize,
                fill=none, 
            	fill opacity=0.2, 
            	draw opacity=1, 
            	text opacity=1, 
                legend columns = 3,
                legend cell align={left},
            	scale = 1,
            },
            legend to name = legend:roc_phd_kf
		]

        \addplot[
   	    color=\colorOne,
        solid,
        mark options={solid},
        line width=1pt,
        forget plot]
    	plot table[x expr=\thisrowno{0}, y expr=1- (\thisrowno{1}+\thisrowno{2})/2] {Data/PHD/\filenamePhd};

        \addplot[
   	    color=\colorTwo,
        solid,
        mark options={solid},
        line width=1pt,
        forget plot]
    	plot table[x expr=\thisrowno{0}, y expr=1- (\thisrowno{1}+\thisrowno{2})/2] {Data/PHD/\filenamePhdOne};

        \addplot[
   	    color=\colorThree,
        solid,
        mark options={solid},
        line width=1pt,
        forget plot]
    	plot table[x expr=\thisrowno{0}, y expr=1- (\thisrowno{1}+\thisrowno{2})/2] {Data/PHD/\filenamePhdPointEight};

        \addplot[
   	    color=\colorOne,
        dashed,
        mark options={solid},
        line width=1pt,
        forget plot]
    	plot table[x expr=\thisrowno{0}, y expr=1 - ((\thisrowno{1}+\thisrowno{1})/2)] {Data/kf_roc/kf_roc_300ms.txt};

        \addplot[
   	    color=\colorTwo,
        dashed,
        mark options={solid},
        line width=1pt,
        forget plot]
    	plot table[x expr=\thisrowno{0}, y expr=1 - ((\thisrowno{1}+\thisrowno{2})/2)] {Data/kf_roc/kf_roc_100ms.txt};
        
        \addplot[
   	    color=\colorThree,
        dashed,
        mark options={solid},
        line width=1pt,
        forget plot]
    	plot table[x expr=\thisrowno{0}, y expr=1 - ((\thisrowno{1}+\thisrowno{2})/2)] {Data/kf_roc/kf_roc_80ms.txt};
        
        \addlegendimage{Gray, solid, line width=1.2pt}
        \addlegendentry{PHD}

        \addlegendimage{Gray, dashed, line width=1.2pt}
        \addlegendentry{KF}

        \addlegendimage{empty legend}
        \addlegendentry{}
        
        \addlegendimage{\colorOne, solid, line width=1.2pt}
        \addlegendentry{$\tilde{T}_m = \qty{300}{\milli\s}$}

        \addlegendimage{\colorTwo, solid, line width=1.2pt}
        \addlegendentry{$\tilde{T}_m = \qty{100}{\milli\s}$}

        \addlegendimage{\colorThree, solid, line width=1.2pt}
        \addlegendentry{$\tilde{T}_m = \qty{80}{\milli\s}$}

        \end{loglogaxis}
\end{tikzpicture}
    \vspace{-0.5cm}
    \caption{Detection performance for varying observation durations \mbox{$\tilde{T}_m \in {80, 100, 300}$~ms} (FN rate vs. FP rate, averaged over running and walking datasets). 
    Solid: PHD, dashed: KF. 
    Longer observation windows reduce missed detections at the same false-alarm rate, with the PHD filter consistently outperforming the KF baseline.
    }
    \label{fig:res_roc_obs_time}
\end{figure}
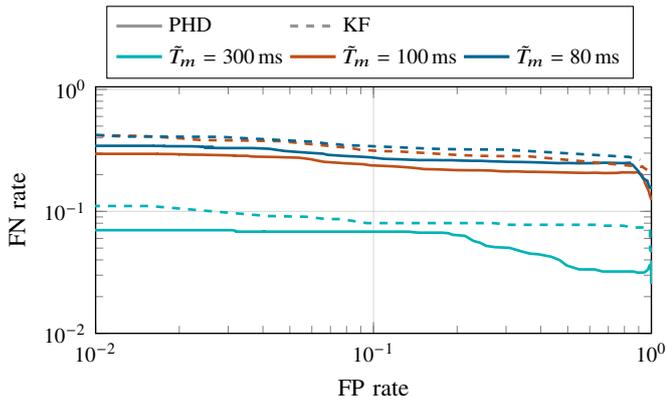

\section{conclusion}
In this paper, we proposed an end-to-end \gls{nlos} intrusion detection system based on a millimeter-wave \gls{isac} \gls{poc}, validated through real-world experiments with moving human targets at the ARENA2036 industrial research campus.
By leveraging reflections off a large surface and applying \gls{kf}- and \gls{phd} filter-based tracking, we demonstrated that reliable detection of moving target presence in a cluttered indoor industrial \gls{nlos} environment is possible without raising false alarms. 
Experiments involving a synthetic \gls{fp} set further emphasize the robustness of our system against false alarms.

In the future, the \gls{nlos} target state estimation can be refined through additional processing steps.
For example, after target detection, advanced features, such as micro-Doppler traces, can be extracted for classification or attitude estimation.
Furthermore, the showcased method should be validated in different scenarios, such as urban intersections, to assess its generalizability.

\if\withAck1
\section{acknowledgments}
Paolo Tosi has been supported by the European Commission through the ISLANDS project (grant agreement no. 101120544).
This work has also received support from the Federal Ministry of Research, Technology and Space of Germany in the project “SENSATION” under grant number 16KIS2523K.
\fi

\bibliographystyle{IEEEtran}
\bibliography{referencing/bibliography}

\end{document}